\documentclass{article}

\usepackage{arxiv}

\usepackage[utf8]{inputenc} 
\usepackage[T1]{fontenc} 
\usepackage{hyperref} 
\usepackage{url} 
\usepackage{booktabs} 
\usepackage{amsfonts} 
\usepackage{nicefrac} 
\usepackage{microtype} 
\usepackage{lipsum}		
\usepackage{graphicx}
\usepackage{doi}

\usepackage[numbers]{natbib}

\title{Assessing the Performance of 1D-Convolution Neural Networks to Predict Concentration of Mixture Components from Raman Spectra}

\author{ \href{https://orcid.org/0000-0001-7181-8270}{\includegraphics[scale=0.06]{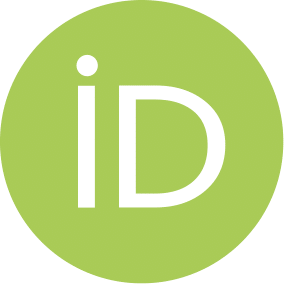}\hspace{1mm}Dexter Antonio}\thanks{\href{https://github.com/dexterantonio)}{GitHub}} \\
	Department of Chemical Engineering\\
	University of California, Davis\\
	Davis, CA\\
	\texttt{ddantonio@ucdavis.edu} \\
	\And
	\href{https://orcid.org/0000-0002-1843-9332}{\includegraphics[scale=0.06]{orcid.png}\hspace{1mm}Hannah O'Toole} \\
	Department of Biomedical Engineering\\
	University of California, Davis\\
	Davis, CA\\
	\texttt{hjotoole@ucdavis.edu} \\
 	\And
	\href{https://orcid.org/0000-0001-8193-1664}{\includegraphics[scale=0.06]{orcid.png}\hspace{1mm}Randy Carney} \\ Department of Biomedical Engineering\\
	University of California, Davis\\
	Davis, CA\\
	\texttt{rcarney@ucdavis.edu} \\
 	\And
 \href{https://orcid.org/0000-0001-9834-8264}{\includegraphics[scale=0.06]{orcid.png}\hspace{1mm}Ambarish Kulkarni } \\
	Department of Chemical Engineering\\
	University of California, Davis\\
	Davis, CA\\\
 	\texttt{arkulkarni@ucdavis.edu}
 \And
 \href{https://orcid.org/0000-0002-4455-8749}{\includegraphics[scale=0.06]{orcid.png}\hspace{1mm}Ahmet Palazoglu} \\
	Department of Chemical Engineering\\
	University of California, Davis\\
	Davis, CA\\
	\texttt{anpalazoglu@ucdavis.edu}}

\renewcommand{\shorttitle}{Assessing the Performance of 1D-Convolution Neural Networks to Predict Concentration of Mixture Components from Raman Spectra}

\hypersetup{
pdftitle={A template for the arxiv style},
pdfsubject={q-bio.NC, q-bio.QM},
pdfauthor={David S.~Hippocampus, Elias D.~Striatum},
pdfkeywords={Raman Spectroscopy, Machine Learning, Mixtures, Bioprocess Control, Signal Processing},
}

\begin{document}
\maketitle

\begin{abstract}
	An emerging application of Raman spectroscopy is monitoring the state of chemical reactors during biologic drug production. Raman shift intensities scale linearly with the concentrations of chemical species and thus can be used to analytically determine real-time concentrations using non-destructive light irradiation in a label-free manner. Chemometric algorithms are used to interpret Raman spectra produced from complex mixtures of bioreactor contents as a reaction evolves. Finding the optimal algorithm for a specific bioreactor environment is challenging due to the lack of freely available Raman mixture datasets. The RaMix Python package addresses this challenge by enabling the generation of synthetic Raman mixture datasets with controllable noise levels to assess the utility of different chemometric algorithm types for real-time monitoring applications. To demonstrate the capabilities of this package and compare the performance of different chemometric algorithms, 48 datasets of simulated spectra were generated using the RaMix Python package. The four tested algorithms include partial least squares regression (PLS), a simple neural network, a simple convolutional neural network (simple CNN), and a 1D convolutional neural network with a ResNet architecture (ResNet). The performance of the PLS and simple CNN model was found to be comparable, with the PLS algorithm slightly outperforming the other models on 83\% of the data sets. The simple CNN model outperforms the other models on large, high noise datasets, demonstrating the superior capability of convolutional neural networks compared to PLS in analyzing noisy spectra. These results demonstrate the promise of CNNs to automatically extract concentration information from unprocessed, noisy spectra, allowing for better process control of industrial drug production. Code for this project is available at github.com/DexterAntonio/RaMix. 
 
\end{abstract}


\section{Introduction}
Industrial fermentation processes are used in the pharmaceutical industry to produce biologic drugs, many of which have therapeutic effects unachievable with traditional small molecule pharmaceuticals \citep{1, 2}. Despite these unique therapeutic benefits, biologic drugs are challenging to characterize and manufacture \citep{3}. Specifically, minute differences in process conditions can subtly alter a protein drug’s structure, such as its glycosylation pattern, and reduce the effectiveness of the drug \citep{3, 4}. To ensure that the critical quality attributes (CQAs) of pharmaceuticals are met, the biopharmaceutical industry has adopted a quality-by-design (QbD) strategy, which achieves high product quality by controlling critical process parameters (CPPs) during the production process \citep{5}. By maintaining CPPs with process control techniques exacting drug specifications can be met. Complementing the QbD are process analytic tools (PAT) that provide accurate measurements of the production systems state, allowing for more advanced process control and further increasing product quality \citep{6, 7}. 

Real time process analytical tools which minimize or eliminate the lag between sample collection and measurement are needed to detect production anomalies sufficiently early to implement corrective actions \citep{5}. Some system attributes such as dissolved oxygen, pH, and temperature can easily be measured with embedded sensors \citep{8}, whereas others such as product concentrations, and glucose levels cannot \citep{4}. Traditional methods for biochemical composition measurements are liquid chromatography (LC), mass spectrometry (MS), and combination methods (LCMS) \citep{9}. Although these techniques are well developed and highly accurate, they require sample collection, extensive sample preparation unique to the analyte class, and trained experts to interpret the results. These drawbacks combined with the high cost of their use (columns, buffers, labor), limit their integration into large-scale production systems \citep{9}. One emerging class of techniques for monitoring the chemical composition of fermentation processes is Raman spectroscopy \citep{6, 9}. Unlike LCMS, Raman spectroscopy can be configured to measure the chemical composition in situ, without the need for sample preparation \citep{9, 10}. This permits real-time, continuous measurements made a distance, i.e., using non-destructive light irradiation. 

A hurdle to the widespread implementation of Raman spectroscopy for process monitoring is the post-collection analysis required to extract chemical concentrations from Raman signals \citep{11, 12}. Addressing this issue, is the key focus on the Ramix package and this paper. Traditional methods to extract concentration information from Raman spectra require pre-processing procedures (baseline correction, smoothing, and normalization) followed by peak fitting, through a semi-manual curve fitting procedure performed by a trained scientist. Advances in baseline correction algorithms and chemometric algorithms allow for the automation of post-collection parameter extraction, permitting online data analysis \citep{5, 11, 12}. Two major classes of chemometric algorithms for Raman spectroscopy analysis are partial least squares (PLS) \citep{13} and convolutional neural network (CNN) methods \citep{14}. 

To assess the performance of different machine learning (ML) algorithms to predict mixture concentrations from Raman spectra, a standard dataset is needed. As of January 2023, no freely available dataset of Raman spectra of mixtures was available. Naturally, the performance of these algorithms depends on the level of pre-processing and noise in the dataset as well as the dataset size, meaning that variations in the noise levels and baseline can affect the overall model performance.

 This lack of a standard dataset and unquantified effect of the baseline on the model performance motivated the creation of RaMix, a Python package capable of producing datasets of mixture spectra with various degrees of noise. By examining how the model performance varies with different levels of augmentations, the strengths and weaknesses of different model classes can be assessed, without the need for massive experimental undertakings. Most importantly, this package only requires a single, individual component spectrum for each mixture compound to generate the synthetic spectra. Thus, the RaMix package is able to produce synthetic spectra that can be used to assess different chemometric algorithms. The purpose of the RaMix package is help address this lack of a standard dataset  by providing simulated spectra that can be used to understand chemometric algorithms performance.

To demonstrate the use of RaMix and demonstrate how it can be useful for assessing the performance of chemometric models, the RaMix package was utilized to generate 48 distinct datasets from mixtures consisting of the chemical signatures of five compounds, namely ethanol, glucose, water, glycerol, and lactic acid. Cellular attenuation of the Raman signal was not included in this simulation. The ML algorithms assessed were partial least squares (PLS), a simple neural network (simpleNN), a simple convolutional neural network (simpleCNN), and a 1D-Convolutional Neural network with a ResNet architecture (ResNet). 

The major contributions of this work are the development of RaMix, an integrated software package that allows for the generation of datasets of synthetic Raman spectra and the assessment of the effect of noise on chemometric algorithm performance with the RaMix package. We first introduce the structure of the RaMix package, before demonstrating this package’s usefulness in assessing the effect of noise on chemometric model performance. Finally, we present an overview of the implications and use cases of this package and the future work needed to further quantify the effect of noise on chemometric methods.

\section{Experimental}
\label{sec:Overivew of Raman Spectroscopy}
A Raman spectrometer probes a solution and generates a Raman spectrum, which contains information about the concentration and identity of each species in the solution based on their inelastic scattering of incident light \citep{23}. Due to their underlying differences in rotational and vibrational Raman active modes, each molecule has a unique Raman spectrum, which consists of several peaks corresponding to each mode. These peaks  can be modeled as Lorentzian (Equation \ref{equation 3}) in shape and are typically sharp with minimal overlap \citep{23}. The position of the peak in the spectra relates to the energy of the vibrational or rotational mode, and this position can be altered by changes in the chemical environment or bonding. 
\begin{equation}
L(x_i)=\frac{a_0}{1+\frac{x_i-p_0}{w/2}^2}
\label{equation 3}
\end{equation}

In Equation \ref{equation 3}, $L(x_i)$ is the intensity of the signal at wavenumber $x_i$, $a_0$ represents the maximum amplitude, $p_0$ is the peak position and $w$ is the full width at the half maximums.
For spontaneous Raman scattering, a spectra of mixtures components are approximately the sum of the individual component spectra, weighted by their concentration. Deviation from this linear relation occurs when non-covalent interactions between the solute molecules alter the chemical bond strength of the solutes, changing the individual component spectra through  peak broadening and minor peak shifts. Variation in laser power and instrument noise can also lead to small peak shifts \citep{23}. Another significant source of noise in Raman spectra is fluorescence, which can be strong enough to overwhelm the Raman signal and limits the usefulness of conventional Raman spectroscopy \citep{9}.

\begin{figure}[!htb]
	\centering
         \includegraphics[scale=0.4]{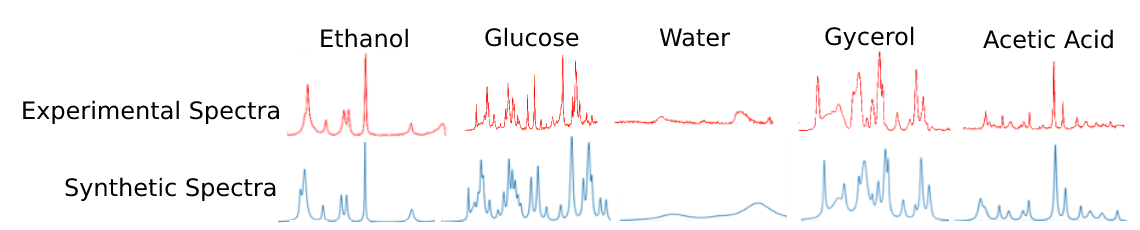}
	\caption{Comparison between the experimental spectra from the Biorad spectra database (red) and the simulated individual component spectra generated from the spectral skeletons (blue). The simulated spectra are generated through the summation of Lorentzian peaks, with parameters manually tuned to ensure resemblance between the experimental and simulated spectra}
	\label{fig:figure1}
\end{figure}

Quantum chemical calculation methods exist for simulating Raman spectra \citep{24} but are computationally expensive. An alternative simulation technique involves using the Lorentzian fitting parameters to recreate the spectral line shapes. This technique reduces the computational expense of simulating spectra while still producing convincing synthetic spectra. Furthermore, by simulating spectra in this manner, a direct understanding of the effect of peak noise on prediction accuracy can be gained. Thus, simulating spectra using Lorentzian line shapes was the technique chosen for the RaMix package. 
\label{sec:Experimental System Inspiration}
An experimental study that utilized Raman spectroscopy for real-time monitoring of ethanol production with baker’s yeast (\emph{saccharomyces cerevisiae}) was chosen as a model for the simulated spectra \citep{10}. In the experimental study, ethanol, glucose, water, glycerol, and lactic acid concentrations were monitored in real-time using a Raman spectrometer embedded in the bioreactor. The data analysis involved using a multi-linear regression model and accurate concentration data from HPLC analysis to relate Raman spectra to sample composition. The analysis was complicated by Raman signal attenuation from high cellular density, thus additional modeling was performed in attempt to compensate for this cell density-related Raman signal attenuation. This study validated the use of Raman spectroscopy to monitor the aforementioned chemical components in a real bioreactor. Therefore in the simulation study the same five chemicals were used (Figure \ref{fig:figure1}). 

The RaMix package requires that the parameters for all the Lorentzian peaks fit to the experimental spectra. These fitting parameters are referred to as “spectral skeletons”. The Bio-Rad Raman spectral database contains freely available images of Raman spectra and was utilized to locate spectra of the five chosen components \citep{25}. Lorentzian peaks were then fitted to these spectra. The spectrum skeletons were visualized by using them to generate Lorentzian peaks and compared to the Biorad spectra in Figure \ref{fig:figure1}. 

\begin{figure}[!htb]
	\centering
         \includegraphics[scale=1]{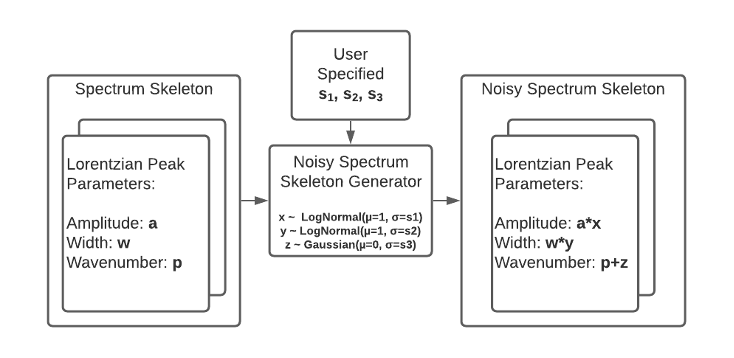}
	\caption{Overview of the noisy spectrum skeleton generator function. The function takes as input a spectrum skeleton and noise parameters and outputs a “noisy spectrum skeleton with noise applied to each peak parameter”.
Random variables sampled from the log-normal distribution were multiplied by the amplitude and the width parameter to generate the noisy spectrum skeleton, where a random variable sampled from the Gaussian distribution was added to the wavenumber position. Preliminary experiments indicated that the noise in the peak location had less of an effect on the prediction compared to the noise in the peak width and amplitude. Thus, in this study, a singular “noise level” was used where $s_1 = noise\ level$ and $s2=s3=noise\ level/10$. By constraining the distribution parameters to a single metric, the effect of noise on spectra prediction could more easily be explored. The noise levels selected were $0$, $0.5$, $1.0$, $1.5$, and $2.0$ to span the somewhat arbitrary qualitative range of “no-noise” to “high noise”.}
	\label{fig:figure2}
\end{figure}

\label{sec:Noise Addition to Spectrum Skeletons }
Noise was added to the individual peak parameters and a random baseline was added to some of the simulated spectra. Noisy spectra skeletons were generated to assess the sensitivity of the different chemometric algorithms to variation in peak location, shape, and height. The noise was added to the spectrum skeleton via a function which input a spectrum skeleton and noise parameters and outputted a noisy spectrum skeleton, with random numbers added and multiplied by each peak parameter (Figure \ref{fig:figure2}).

\begin{figure}[!htb]
	\centering
         \includegraphics[scale=0.15]{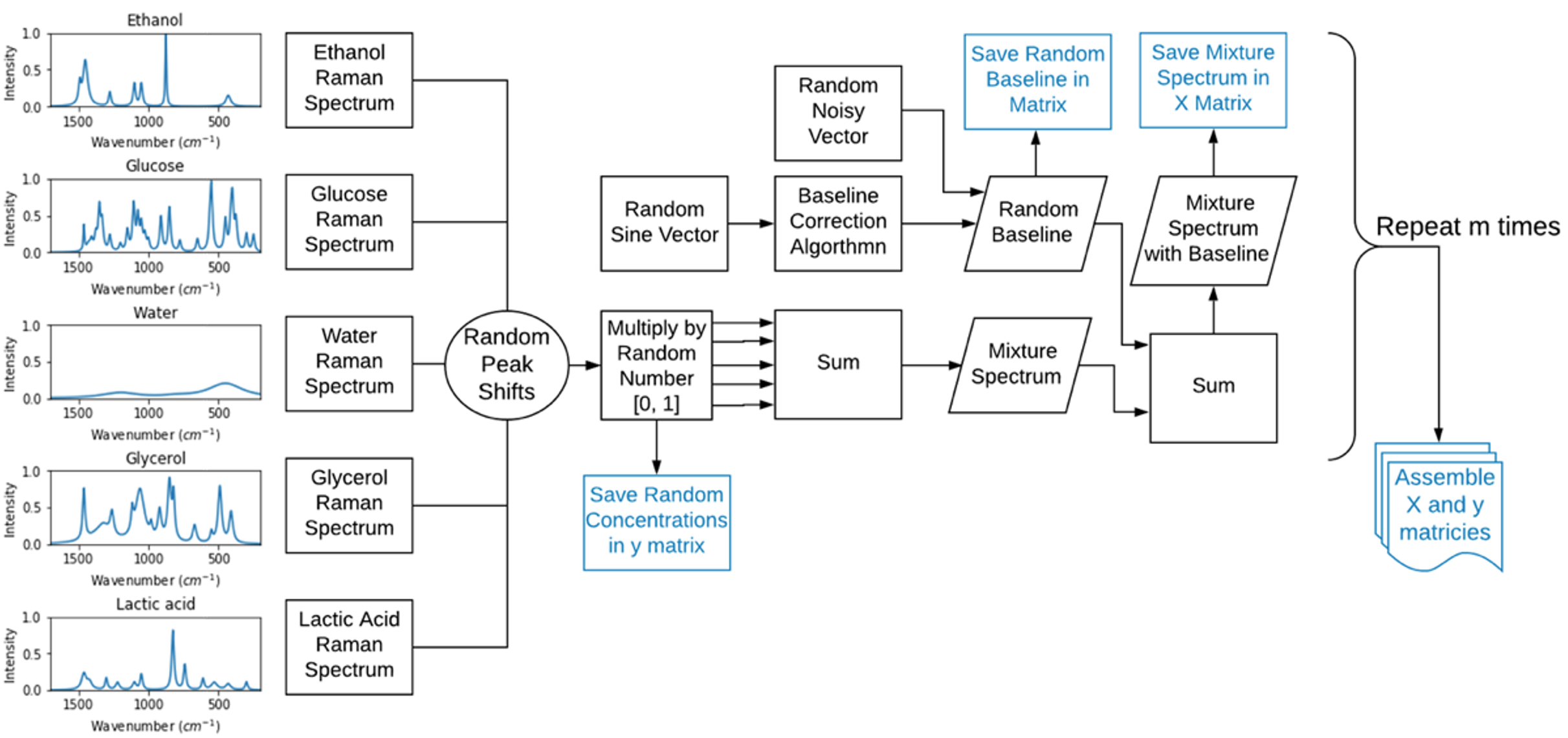}
	\caption{An overview of the random mixture spectra dataset generation. Individual component spectra skeletons are transformed into noisy spectra skeletons. These spectra skeletons are then used to generate simulated individual component spectra, which are randomly scaled and combined to form simulated mixture spectra. This generation process is repeated $m$ times, for each datapoint in a simulated mixture dataset.  The first stage in the creation of the mixture spectrum dataset involved adding noise to each spectrum skeleton to generate noisy spectrum skeletons. After creating the noisy spectrum skeletons, each skeleton was used to generate pure component spectra vectors. The pure component spectra vectors were combined in random linear combinations to generate mixture spectra.}
	\label{fig:figure3}
\end{figure}

\label{sec:Mixture Spectra Dataset Generation}
To assess the performance of the ML algorithms on datasets with different levels of noise, 48 separate datasets were generated. A diagram demonstrating the generation of a single dataset is shown in Figure \ref{fig:figure3}. The formula for the mixture spectra $S_{mixture}$ is defined in equation \ref{mixture_equation}

\begin{equation}
S_{Mixture}= c_1 S_{Ethanol}+c_2 S_{Glucose}+c_3 S_{Water}+c_4 S_{Glycerol}+c_5 S_{Lactic Acid}
\label{mixture_equation}
\end{equation}
In equation \ref{mixture_equation} $c_i$ is a random number between $0$ and $1$ and $S_i$ is a $1x1800$ spectrum row vector.
The weights applied to each spectrum were saved along with the resulting mixture spectra. This process was repeated $m$ times, where $m$ is the size of the datasets ranging from $10$ to $10,000$ for this simulation experiment. 
The range $\left[0, 1\right]$ is not representative of the expected concentration of each species in a fermentation process, but this simplification was made to make the data easier to interpret. Furthermore, the Bio-R ad Raman spectral database contained spectra with normalized peaks, making the intensity of each spectrum arbitrary.
After generating the spectra, additional noise was added to some of the data in the form of a noisy random baseline. This baseline was generated by feeding a noisy random sine vector into a baseline correction algorithm, which tried to fit a polynomial to the sine function \citep{26}. This process generated a random baseline that resembled the fluorescence baselines typically found in Raman spectroscopy, which results from the fluorescent glass composing the bioreactor. Additional random noise was also added to the spectra. The final spectra, baseline, and species concentrations were saved in data matrices, which allowed for the subsequent data analysis to take place.
\label{sec:Data Splitting}
Each dataset was split into a training and testing set with 80\% of the data going to the training set. The training set was used to train the model and estimate the model parameters. By refraining from testing each model on the test set until after the hyperparameter tuning, a more accurate gauge of each model’s performance was obtained.
\label{sec:Chemometric Algorithms}
Four different algorithms were used to predict concentrations from the mixture spectrum: Partial Least Squares (PLS), a simple Artificial Neural Network (simpleNN), a simple Convolution Neural Networks (simpleCNN) and a 1D implementation of ResNet-34 (ResNet). The Python package scikit-learn PLS algorithm was used for the PLS regression \citep{27}. Python’s Tensorflow 2.0 was used to build neural networks and convolution neural networks \citep{28}.
\label{sec:Fitting}

As noted, the 48 datasets were each split into training and test sets. The training set was utilized for fitting and the test set was utilized for model validation. The models were trained on each datasets and their performance was assessed. The spectral datasets were not standardized, however the y matrices were standardized using Equation \ref{ynorm}. 

\begin{equation}
    y_{norm}=\frac{y_{data}-mean(y_{data})}{std(y_{data})}
    \label{ynorm}
\end{equation}

All three models were fitted to the training data set. The NNs and CNNs were trained with a maximum of 200 epochs with early stopping implemented with a 0.75\% training validation split. This early stopping criterion reduced the training data available to the CNN models by 25\% but reduced overfitting by stopping the models before overfitting could occur. 
\label{sec:Metrics}
After the four models were trained on the training data their predictive performance was assessed on the test sets. The three related metrics that were used to assess the model’s performance were explained variance, root mean squared error ($RMSE$), and coefficient of variation ($R^2$). 

\section{Results and Discussion}
\begin{figure}[!htb]
	\centering
         \includegraphics[scale=1]{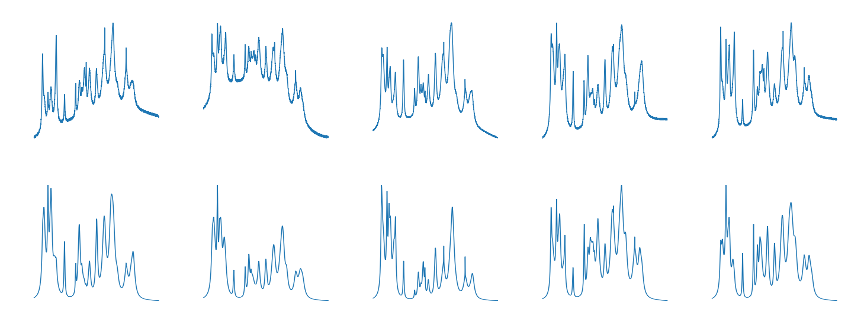}
	\caption{Example simulated Raman spectra with noise levels of $1.5$ with and without baselines with varying species concentrations. The spectra of mixture spectra share approximately common peaks but vary in overall shape, intensity and exact peak location. }
	\label{fig:figure4}
\end{figure}

The purpose of the RaMix package is to provide simulated spectra that can be used to assess the ability of various chemometric algorithms to predict the concentration of chemical species from simulated spectra with varying degrees of noise. To demonstrate this capability, four separate chemometric algorithms were assessed on 48 distinct datasets generated with the RaMix package. The noise level, which is a single metric that encapsulates the noise added to the peak parameters in each spectrum, varied between $0.0$ (no noise) and $2.0$ (significant noise). Example simulated spectra with and without baselines for the noise level $1.5$ are shown in Figure \ref{fig:figure4}.

\begin{figure}[!htb]
	\centering
         \includegraphics[scale=1]{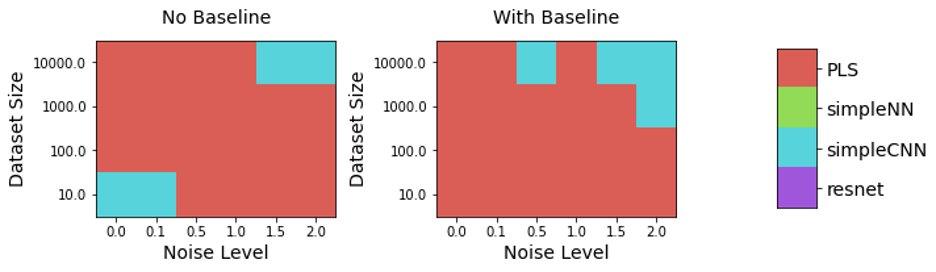}
	\caption{Comparison of model performance on 48 datasets. The model with the highest coefficient of determination, $R^2$, for a given dataset when trained and tested on partitions of the dataset is indicated by colored blocks.}
	\label{fig:figure5}
\end{figure}

The assessment of four chemometric algorithms was performed using the performance metrics, $RMSE$, explained variance, and the coefficient of determination, $R^2$, for each model on each dataset. The $R^2$ value was chosen to select the best performing model and the result of that comparison is shown in Figure \ref{fig:figure5}.

Figure \ref{fig:figure5} shows that the PLS algorithm performed the best for most datasets (40/48) yet was outperformed by the simple CNN model on the higher noise datasets with $10,000$ data points. The strong performance of the PLS algorithm is due to its ability to generate reasonable basis vectors from a small dataset and the inherent assumption about a linear relationship between spectral intensity and species concentration. 

By contrast, CNNs require “large” datasets to fully train the model, make no assumptions about the underlying functional relationship between the input and output vectors, and incorporate the spatial relationship between data points into the model. The additional complexity of the CNN models worsens their performance on small, low-noise datasets without baselines, yet this flexibility allows it to outperform the PLS model on the larger, high noise datasets. The CNN’s convolutional layers also allow for partial baseline removal with trained convolutional layers, explaining the algorithm’s better performance on the large dataset, lower noise spectra with baselines. 

\begin{figure}[!htb]
	\centering
         \includegraphics[scale=1]{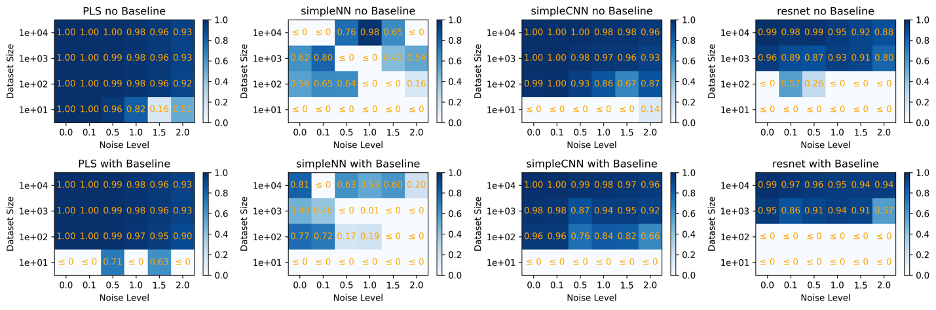}
	\caption{Individual model performance ($R^2$ value) for each dataset rounded to two decimal places. Dark blue squares indicate higher performance and white squares indicate lower performance. In situations where the model did not converge, the R2 value is less than or equal to zero and is indicated by the $\leq 0$ annotation.}
	\label{fig:figure6}
\end{figure}

Another useful comparison is the difference in individual model performance on each dataset. Figure \ref{fig:figure6} shows this performance by displaying the $R^2$ value for each dataset, rounded to two decimal places. 
 
\begin{figure}[!htb]
	\centering
         \includegraphics[scale=1]{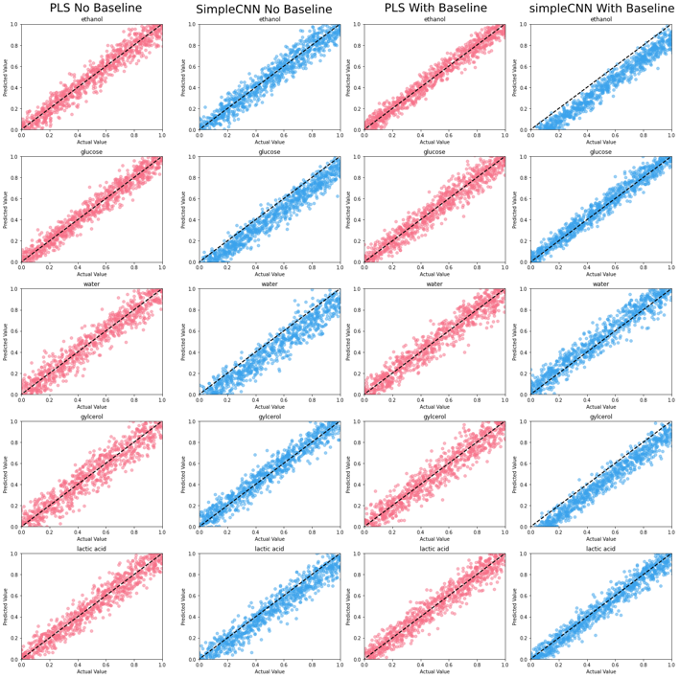}
	\caption{Comparison of the PLS, SimpleNN, simpleCNN resnet algorithm performance on a $10,000$ point dataset with a noise level of $0.5$ with and without a baseline.}
	\label{fig:figure7}
\end{figure}

Figure \ref{fig:figure6} shows the performance for the different models on various datasets. The simpleNN is a poor-performing algorithm, achieving an $R^2$ value below $0.80$ on all but one dataset. The ResNet model does not perform well on the datasets with less than $1,000$ data points, including the low noise datasets. The ResNet architecture was selected because previous studies that used CNNs for Raman spectra classification tasks achieved high performance with ResNet architectures. The poor performance can be explained by noting that the ResNet architecture has many more parameters to train than the simple CNN. 
Figure 9 also reveals the similar performance between the PLS and simpleCNN algorithm on datasets with $1,000$ and $10,000$ data points in the low noise categories ($0.0$ and $0.1$), with each model achieving comparably high performance ($R^2 > 0.98$). The PLS algorithm is insensitive to the dataset size above $100$, yet the CNN algorithm continues to improve with increasing dataset size indicating that the performance of the CNN algorithm has more potential for improvement.
To further compare the PLS and simpleCNN models, their performance on two datasets was examined in detail. The datasets selected for this analysis were the largest ($10,000$ data points) datasets with a noise level of $0.5$, one having a baseline and the other without an added baseline. A calibration plot for the two best-performing models on these datasets (PLS and simpleCNN) is shown in Figure \ref{fig:figure7}. 

Figure \ref{fig:figure7} indicates that, for the PLS algorithm, the predictions are evenly clustered around the $x=y$ line, demonstrating that model is not biased. In contrast, the predictions for glycerol in the simpleCNN with the baseline is mostly under the $x=y$ line indicating a chronic underprediction of the glycerol concentration. Surprisingly, neither model is dramatically affected by the addition of a baseline, with its presence only leading to a slightly more dispersed clustering around the $x=y$ line. The predicted versus actual plots also verify the good performance of both models in predicting concentration data from the simulated spectra and show the potential of CNNs to act as a substitute for the PLS algorithm. 

One of the main shortcomings of this simulation experiment is that the linear way in which the data is generated does not account for spectra changes that result from interactions among the solutes. Another shortcoming is that the cellular attenuation of the Raman signal, an important confounder in real-time process monitoring, is not incorporated into this model \citep{10}. The performance of the PLS algorithm was bolstered by the simple manner with which the data was generated and by ignoring cellular attenuation. By adding these sources of nonlinear behavior into the data generation step, the CNN performance would likely surpass that of the PLS algorithm. Nevertheless, the RaMix model provides an important first step in generating datasets useful for comparing the performance of chemometric algorithms and can easily be extended to other systems. 

\section{Conclusions}
The RaMix package allows for the generation of synthetic Raman spectra, which can be used to assess the performance of chemometric algorithms to predict concentrations from simulated spectra. In this study, $48$ datasets were generated with the package, and the predictive ability of CNNs and PLS algorithms was assessed. CNNs were shown to have comparable performance to the standard PLS algorithm and outperformed the PLS algorithm in certain high-noise situations. 
The lack of large public Raman spectroscopy datasets makes validating that the conclusions drawn from the synthetic data with experimental data challenging. Nevertheless, the release of this package is an important first step in the generation of standard datasets for chemometric model assessment and will hopefully catalyze the public release of the large spectra datasets needed for chemometric model refinement. Synthetic data also provides the ability to probe the effects of distinct types of noise on chemometric model performance and can complement the conclusions found from the analysis of real experimental spectra. 
In addition, the use of convolutional neural networks for multi-component mixture concentration predictions has not previously been reported, and thus this study shows a new application of CNNs to address a pressing issue in biotechnology. Specifically, these results demonstrate that the use of CNNs for the prediction of mixture concentrations holds considerable promise for the automated real-time monitoring of bioreactors. 
\section{Code Availability}
Code for this project is available at \href{github.com/DexterAntonio/RaMix}{github.com/DexterAntonio/RaMix} under an MIT license.

\bibliographystyle{unsrt}  
\bibliography{references}  

\begin{thebibliography}{10}

\bibitem{1}
Patrick Chames, Marc Van~Regenmortel, Etienne Weiss, and Daniel Baty.
\newblock Therapeutic antibodies: successes, limitations and hopes for the
  future: Therapeutic antibodies: an update.
\newblock 157(2):220--233.

\bibitem{2}
Henrique Neves and Hang~Fai Kwok.
\newblock Recent advances in the field of anti-cancer immunotherapy.
\newblock 3:280--288.

\bibitem{3}
Arnold~G. Vulto and Orlando~A. Jaquez.
\newblock The process defines the product: what really matters in biosimilar
  design and production?
\newblock 56:iv14--iv29.

\bibitem{4}
Peiqing Zhang, Susanto Woen, Tianhua Wang, Brian Liau, Sophie Zhao, Chen Chen,
  Yuansheng Yang, Zhiwei Song, Mark~R. Wormald, Chuanfei Yu, and Pauline~M.
  Rudd.
\newblock Challenges of glycosylation analysis and control: an integrated
  approach to producing optimal and consistent therapeutic drugs.
\newblock 21(5):740--765.

\bibitem{5}
André Guerra, Moritz Stosch, and Jarka Glassey.
\newblock Toward biotherapeutic product real-time quality monitoring.
\newblock 39(3):289--305.

\bibitem{6}
Karen~A. Esmonde-White, Maryann Cuellar, Carsten Uerpmann, Bruno Lenain, and
  Ian~R. Lewis.
\newblock Raman spectroscopy as a process analytical technology for
  pharmaceutical manufacturing and bioprocessing.
\newblock 409(3):637--649.

\bibitem{7}
Moo~Sun Hong, Kristen~A. Severson, Mo~Jiang, Amos~E. Lu, J.~Christopher Love,
  and Richard~D. Braatz.
\newblock Challenges and opportunities in biopharmaceutical manufacturing
  control.
\newblock 110:106--114.

\bibitem{8}
Baowei Wang, Zhiwen Wang, Tao Chen, and Xueming Zhao.
\newblock Development of novel bioreactor control systems based on smart
  sensors and actuators.
\newblock 8:7.

\bibitem{9}
Jens~A. Iversen, Rolf~W. Berg, and Birgitte~K. Ahring.
\newblock Quantitative monitoring of yeast fermentation using raman
  spectroscopy.
\newblock 406(20):4911--4919.

\bibitem{10}
Shuxia Guo, Thomas Bocklitz, and Jürgen Popp.
\newblock Optimization of raman-spectrum baseline correction in biological
  application.
\newblock 141(8):2396--2404.

\bibitem{11}
Shuxia Guo, Petra Rösch, Jürgen Popp, and Thomas Bocklitz.
\newblock Modified {PCA} and {PLS}: Towards a better classification in raman
  spectroscopy–based biological applications.
\newblock 34(4):1--10.

\bibitem{12}
Jinchao Liu, Margarita Osadchy, Lorna Ashton, Michael Foster, Christopher~J.
  Solomon, and Stuart~J. Gibson.
\newblock Deep convolutional neural networks for raman spectrum recognition: A
  unified solution.

\bibitem{13}
Cross decomposition — scikit-learn 0.24.2 documentation.

\bibitem{14}
Yasir Aslam and Santhi N.
\newblock A review of deep learning approaches for image analysis.
\newblock In {\em 2019 International Conference on Smart Systems and Inventive
  Technology ({ICSSIT})}, pages 709--714. {IEEE}.

\bibitem{23}
F.~Pedregosa, G.~Varoquaux, A.~Gramfort, V.~Michel, B.~Thirion, O.~Grisel,
  M.~Blondel, P.~Prettenhofer, R.~Weiss, V.~Dubourg, J.~Vanderplas, A.~Passos,
  D.~Cournapeau, M.~Brucher, M.~Perrot, and E.~Duchesnay.
\newblock Scikit-learn: Machine learning in {P}ython.
\newblock {\em Journal of Machine Learning Research}, 12:2825--2830, 2011.

\bibitem{24}
Martín Abadi, Ashish Agarwal, Paul Barham, Eugene Brevdo, Zhifeng Chen, Craig
  Citro, Greg Corrado, Andy Davis, Jeffrey Dean, Matthieu Devin, Sanjay
  Ghemawat, Ian Goodfellow, Andrew Harp, Geoffrey Irving, Michael Isard,
  Yangqing Jia, Rafal Jozefowicz, Lukasz Kaiser, Manjunath Kudlur, Josh
  Levenberg, Dan Mané, Rajat Monga, Sherry Moore, Derek Murray, Chris Olah,
  Mike Schuster, Jonathon Shlens, Benoit Steiner, Ilya Sutskever, Kunal Talwar,
  Paul Tucker, Vincent Vanhoucke, Vijay Vasudevan, Fernanda Viégas, Oriol
  Vinyals, Pete Warden, Martin Wattenberg, Martin Wicke, Yuan Yu, and Xiaoqiang
  Zheng.
\newblock Tensorflow: Large-scale machine learning on heterogeneous distributed
  systems, 2015.

\bibitem{25}
Kevin Buckley and Alan~G. Ryder.
\newblock Applications of raman spectroscopy in biopharmaceutical
  manufacturing: A short review.
\newblock 71(6):1085--1116.
\newblock Publisher: {SAGE} Publications Ltd {STM}.

\bibitem{26}
Michael S.
\newblock Lineshapes in {IR} and raman spectroscopy: A primer.
\newblock 30(11):42--46.
\newblock Publisher: {MJH} Life Sciences.

\bibitem{27}
Shixuan He, Wei Zhang, Lijuan Liu, Yu~Huang, Jiming He, Wanyi Xie, Peng Wu, and
  Chunlei Du.
\newblock Baseline correction for raman spectra using an improved asymmetric
  least squares method.
\newblock 6(12):4402--4407.
\newblock Publisher: The Royal Society of Chemistry.

\bibitem{28}
Zhi-Min Zhang, Shan Chen, and Yi-Zeng Liang.
\newblock Baseline correction using adaptive iteratively reweighted penalized
  least squares.
\newblock 135(5):1138--1146.
\newblock Publisher: The Royal Society of Chemistry.

\end{thebibliography}


\end{document}